\documentclass[%
 reprint,
 amsmath,amssymb,
 aps,
]{revtex4-2}

\usepackage{graphicx}
\usepackage{dcolumn}
\usepackage{bm}
\usepackage{placeins}
\usepackage{float}
\usepackage{svg}
\usepackage{afterpage}

\begin{document}

\preprint{APS/123-QED}

\title{Electron-beam annealing of Josephson junctions for frequency tuning of quantum processors}

\author{Yashwanth Balaji}
\author{Narendra Acharya}
\author{Robert Armstrong}
\author{Kevin~G.~Crawford}
\author{Sergey Danilin}
\author{Thomas Dixon}
\author{Oscar~W.~Kennedy}
\author{Renuka Devi Pothuraju}
\author{Kowsar Shahbazi}
\author{Connor~D.~Shelly}\email{cshelly@oxfordquantumcircuits.com}

\affiliation{%
 Oxford Quantum Circuits, Thames Valley Science Park, Shinfield, Reading, United Kingdom, RG2 9LH}%

\date{\today}

\begin{abstract}
Superconducting qubits are a promising route to achieving large-scale quantum computers. A key challenge in realising large-scale superconducting quantum processors involves mitigating frequency collisions. In this paper, we present an approach to tuning fixed-frequency qubits with the use of an electron beam to locally anneal the Josephson junction. The technique shows an improvement in wafer scale frequency targetting by assessing the frequency collisions in our qubit architecture. Coherence measurements are also done to evaluate the performance before and after tuning.  The tuning process utilises a standard electron beam lithography system, ensuring  reproducibility and implementation by any group capable of fabricating these Josephson junctions. This technique has the potential to significantly improve the performance of large-scale quantum computing systems, thereby paving the way for the future of quantum computing.

\end{abstract}

\maketitle


\section{\label{sec:level1}Introduction}

Quantum computing research is increasingly focused on developing large-scale and robust quantum processors capable of delivering reliable computation for real-world applications \cite{Martinis2015}. The pursuit of fault-tolerant quantum computers with effective error mitigation resulted in recent demonstrations showcasing a quantum advantage over classical computers \cite{Arute2019,Bravyi2023,Kim2023}. This advancement is facilitated by the growing number of qubits making up the quantum processing units (QPUs) and the performance of those qubits, with key metrics being high coherence time, high one- and two-qubit gate fidelity, and minimal cross talk between qubits \cite{Hertzberg2020}.

Superconducting qubits, a leading and widely used platform for universal gate-based quantum computing rely on Josephson junctions (JJs), a non-linear inductive element to provide the anharmonic energy spacing required for qubit operation. Fixed-frequency transmons have demonstrated gate fidelities surpassing 99\,\%  \cite{Hong2020,Chen2023,Marxer2023}. In order to achieve such high fidelity at scale, it is necessary to ensure accurate targeting of qubit frequencies \cite{Hertzberg2020,Kreikebaum_2020} and optimised frequency allocation \cite{Hertzberg2020,Kreikebaum_2020,8614500,Morvan2022} to avoid a series of frequency collisions between the first, and higher order transitions of neighbouring qubits.  In contrast to flux tunable qubit designs, fixed-frequency transmons are more sensitive to fabrication imperfections, translating to inaccuracies in target frequencies and frequency collisions between qubits with no active control to mitigate them.

In the transmon regime, the qubit frequency, $f_{01}\simeq(\sqrt{8{E_{J}E_{C}\vphantom{E^2}}}-E_{C})/h$, where the Josephson energy, $E_{J}={\hbar I_c}/{{2e}}$ is much greater than the charging energy, $E_{C}={e^{2}}/{{2C}}$ \cite{Koch2007}, where ${C}$ corresponds to the transmon capacitance and $I_{c}$  is the critical current provided by  the Ambegaokar-Baratoff relation $I_{c}={\pi\Delta}/{2eR_{\mathrm{n}}}$ \cite{AB-relation}. $R_{\mathrm{n}}$ corresponds to the resistance of the tunnelling barrier, and ${\Delta}$, the superconducting gap. A major challenge lies in achieving precise control of  ${R_n}$  across multiple JJs in a QPU,  which is greatly influenced by current fabrication methods. This variability in $R_{\mathrm{n}}$ directly translates to variations in $hf_{01}$ and thus  impacts the ability to accurately and precisely allocate qubit frequencies.

\begin{figure*}
\includegraphics[width=1.05\textwidth]{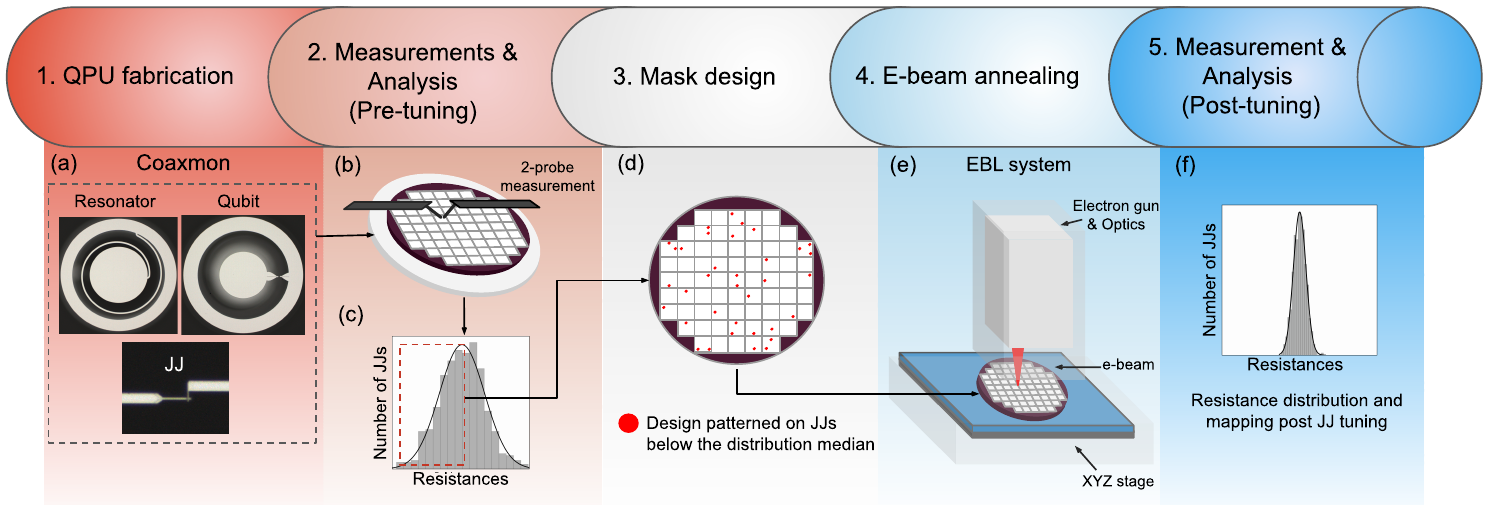}
\caption{\label{fig:procedure}Procedure for e-beam tuning of qubits. 1. Qubits or QPUs fabricated at wafer scale with specific frequencies targeted by varying the JJ dimension. (a) shows the coaxmon design of Resonators, Qubit and JJs. 2. Measurement of the junction barrier resistance $R_{\mathrm{n}}$ at room temperature performed across the wafer shown by (b). (c) Key parameters obtained such as wafer-scale, and QPU-scale resistance spread $\sigma_{\mathrm{R}}$. Further data analysis performed to identify qubits that require e-beam tuning to optimise frequency targeting. 3. (d) Qubit-specific lithography mask created in order to perform junction tuning on selected JJs only. 4 (e) Electron-beam annealing performed on all target junctions. Doses and/or proximity to JJ may vary. 5. (f) Measurements of $R_{\mathrm{n}}$ at room temperature are again performed and post-tuning parameters of spread are determined.}
\end{figure*}

The variability in $R_{\mathrm{n}}$ arises from the Al/AlO$_{x}$/Al stack in the JJs, that are commonly fabricated using double-angled shadow evaporation. Current state-of-the-art shadow-evaporated junctions exhibit a resistance spread between, $\sigma_{\mathrm{R}}$ = 2\,\% -  3\,\% at wafer scale (2 and 3 inch) depending on the JJ dimension \cite{Kreikebaum_2020,8614500,Moskalev2023,Osman2023}. Both lateral dimensions of the JJ and the tunneling barrier itself impacts this resistance spread. Dimensional non-uniformities include line edge roughness, and resist stack inhomogeneity \cite{Pishchimova2023}, resist residues \cite{Quintana2014}, uneven distribution of metal during deposition \cite{Moskalev2023} and variation in critical feature size inherent to tolerance of the lithography tool itself. The tunnelling barrier dimensions and stoichiometry are influenced by the oxidation conditions \cite{Nik2016,Fritz2019}, non-uniform oxide growth across the junction, and several others \cite{Zeng2016,Lapham2022,Bayros2023}. To further scale quantum processors, individual qubit frequency targetting must be improved to ensure high gate fidelities, the requirements for this frequency targetting exceeds what is possible with current state-of-the-art as-fabricated Josephson junctions \cite{Hertzberg2020, Zhang2022}.

To address this as-fabricated variability in $R_{\mathrm{n}}$, a technique to individually tune the JJs post-fabrication is required. Techniques to address this include the use of a laser to fine tune qubit frequencies through localised annealing of JJs \cite{Hertzberg2020,VanderMeer2021}  and the use of repeated DC current pulses through the Josephson junction \cite{Pappas2024}. In this paper, we propose an alternative approach using an electron beam (e-beam) to locally anneal the JJs and thereby reduce $R_{\mathrm{n}}$ variability \cite{EBLA-patent}. This technique takes advantage of a controllable e-beam such as that in an e-beam lithography system (EBL) to individually anneal JJs and thus tune qubit frequency.

Electron beam lithography systems widely available in nanofabrication facilities allow this technique to be easily intergrated into fabrication workflows. The technique shows minimal yield reduction, and minimal impact to the qubit coherence metrics.  We perform wafer level tuning on OQC's Lucy-generation 8QPU and achieve a resistance spread, $\sigma_{R} <  $ 1.5\,\%. Further, we assess the QPU viability by calculating the frequency-induced collisions before and after e-beam tuning. Lastly, we present the results of qubit coherence measurements showing that the qubits retain their high-coherence after electron-beam annealing.

\section{\label{sec:level2}Resistance tuning mechanism and procedure using e-beam}

The barrier resistance of a JJ has been shown to increase over time, in a process commonly referred to as `ageing'. This is an important consideration when fabricating a QPU with precise qubit frequency targets, as those frequencies may drift.  The process of JJ ageing appears related to oxygen diffusion or structural alterations in the AlO$_{x}$ layer that can be accelerated under elevated temperatures \cite{Alexey2011}.  We leverage this ageing process in order to address the variability in JJ resistances post fabrication.

The heating process of an electron beam differs significantly from that of a laser source \cite{Kimmitt1981}.  When a high-energy electron beam interacts with a substrate, the primary electrons of the beam undergo scattering events, both on the surface and within the underlying substrate generating secondary effects, such as secondary electrons, backscattered electrons, and Auger electrons. These events lead to a rise in the localised temperature of the irradiated material \cite{Groves1996, Jiang2015}. 

The scattering mechanisms comprise a combination of both elastic and inelastic scattering. Elastic scattering pertains to the interaction between incoming electrons and atoms, involving the conservation of  kinetic energy and momentum transfer from an energetic electron to an atom. This interaction can result in a collision and direct atomic displacement also known as a knock-on effect. However, we do not expect any knock-on damage as the energy required to displace atoms significantly exceeds the incident beam energies being used \cite{Jiang2015,Egerton2004}. In contrast, inelastic scattering is predominantly an electron-electron interaction wherein a substantial amount of energy is transferred during the process, resulting in electron excitation and ionisation of lattice atoms. A portion of this energy is converted into thermal energy within the substrate, causing temperature increase \cite{Egerton2004, Park2021}.

The parameters for heating a substrate using an e-beam are influenced by several factors including the beam energy, the spot size, current of the electron beam and the time of irradiation \cite{Groves1996, Wang2019}. Therefore, quantifying the amount of heat generated is challenging as e-beam heating behaviour differs significantly between insulating and conducting materials with different densities and thermal conductivities  \cite{Jiang2015}. Previous studies implemented the use of thin film thermocouples  and  nano-watt calorimeters to directly quantify heat production when subjecting materials to an e-beam using transmission electron microscopy (TEM) and scanning electron microscopy (SEM) \cite{Wang2019, Park2021}. Investigations revealed that e-beam irradiation can cause materials to reach a wide rage of temperatures between 10~K and 1000~K \cite{Wang2019}.  Therefore, experiments involving e-beam annealing must be tailored to the specific material stack under irradiation. To achieve this, we have used an EBL system to carry out a series of experiments involving different doses and beam currents on the JJs while maintaining a fixed acceleration voltage. The EBL system not only allows us the capability to modify both the size and shape of the irradiated area but also provides the flexibility to vary the exposure scan pattern, and can be operated in either continuous or pulsed modes.

Figure \ref{fig:procedure} illustrates the step-by-step procedure to tune the JJ resistance. 1) The process begins with the fabrication of either JJ test structures or wafer-scale QPU's with lumped elements  (resonators and capacitors) fabricated using photolithography, followed by JJs fabricated using EBL and a double angled shadow evaporation technique \cite{Rahamim2017}. Figure \ref{fig:procedure}a shows OQC's double-sided coaxmon design of the resonator, qubit and JJ elements. 2) We then conduct room temperature resistance measurements on the JJ tunnel barrier to evaluate their resistance distribution $\sigma_{\mathrm{R}}$, target and yield (Figure \ref{fig:procedure}b and c). Additionally for QPUs, we also derive qubit parameters, such as assessing any potential frequency collisions and calculating the instances where they may occur within the QPUs. To demonstrate a reduction in resistance spread with this technique,  we select JJs in the histogram with resistances below the median for tuning, aiming to increase their resistance and narrow the overall distribution (highlighted by the red dashed line in Figure \ref{fig:procedure}c.) 3) These specific JJs are identified and mapped within the sample design layout (Figure \ref{fig:procedure}d) and were assigned the same dose such that the JJs shift as an ensemble to narrow the spread. A more effective approach would be to assign appropriate e-beam doses  based on their respective positions in the distribution. 4) Figure \ref{fig:procedure}e shows the tuning being carried out in an EBL system, where the designated JJs are exposed. As the qubits used here are not grounded we use a charge mitigation layer in this processing step to avoid charge build up and any associated yield-reducing processes. 5) After tuning, the resistance measurements are performed again to evaluate the tuned resistance distribution, $\sigma_{\mathrm{Re}}$, where R$_{\mathrm{e}}$ corresponds to the e-beam annealed resistance, and assess any change in frequency collision metrics. Based on the results, the QPU's can either be used in a quantum computer, or will undergo another iteration of tuning to further narrow  $\sigma_{\mathrm{Re}}$.

\begin{figure}
\includegraphics[width=0.4\textwidth]{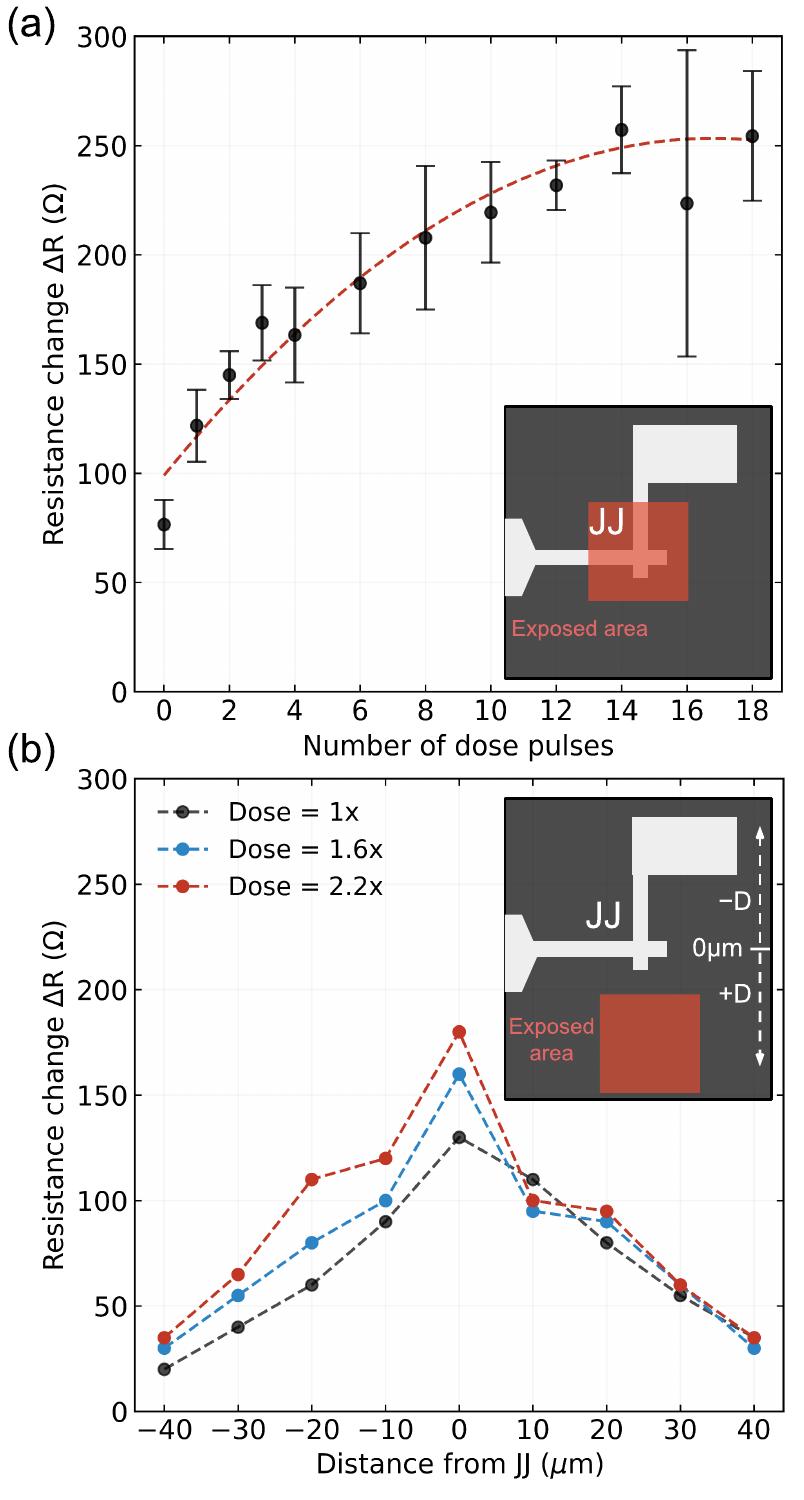}
\caption{\label{fig:dose}(a) Resistance tuning (in absolute $\Omega$) as a function of the number of shots applied to the junction area. Each point represents the average of 16 JJs, with the standard deviation given by the error bars and the red dashed line serving as a visual guide. The inset shows the location of the exposed area. By varying the number of dose pulses that the EBL tool applies to the junction, the absolute resistance change, $\Delta{R}$, will vary. This allows the user to tune the resistance on a per-junction basis. (b) It is also possible to vary the resistance change that the junction undergoes by tuning the proximity of the electron beam to the junction.  The inset shows the schematic of the experiment with distance D placed both above (-D), and below (+D) the JJ. The distance from the JJ is defined from the center of the square exposed area. The exposure area is 15 $\mu\mathrm{m}\times$ 15 $\mu\mathrm{m}$}
\end{figure}

\section{\label{sec:level5}Resistance tuning with applied dose and proximity}

We investigate the influence of dose and proximity of the e-beam on tuning  $R_{\mathrm{n}}$. The operating acceleration voltage of the EBL is kept at 100 keV for all experiments in this work. The study comprises of JJ test structures fabricated with identical dimensions which were exposed with an increasing number of pulses for a fixed e-beam dose.  Figure \ref{fig:dose}a shows the JJ resistance change, $\Delta$R, with increasing number of dose pulses when the exposed area is incident on the JJ, as shown in the inset.  We observe an increase in  $\Delta$R  which saturates at  $\Delta$R = 250~$\Omega$, which corresponds to a resistance tuning of approximately 3\,\%. 
We observe that  the JJs without e-beam exposure on the same sample also show some resistance increase. This is attributed to the relative normal ageing that occurs during the time between initial JJ fabrication and their subsequent e-beam annealing and measurement. The magnitude of resistance change for these unexposed junctions agree with expected ambient ageing in relevant timescales \cite{Koppinen2007}.
          
We next investigate the impact of annealing with respect to the e-beam placement relative to the JJ, as depicted in the inset of Figure \ref{fig:dose}b. The distance, (D), of the e-beam exposure is measured either from above (+D) or below (-D) the JJ from the JJ center. Figure \ref{fig:dose}b  shows the resistance tuning with respect to the beam distance for three different doses. The results show that the maximum resistance
tuning occurs when the e-beam is positioned directly over the JJ. The amount of tuning reduces as the exposed area is moved away from the junction. Our QPU design has a qubit pitch much larger than $40\,\mu\mathrm{m}$, therefore when annealing is performed on the target JJ, none of the neighbouring JJs are affected. These results demonstrate our capability to tune JJ resistance based upon the principle of localised heating of the substrate. Controlling proximity of heating relative to the JJ not only allows greater control of tuning, but also avoids any unwanted tuning of neighbouring qubits based on distance and separation.

\begin{figure}
\includegraphics[width=0.4\textwidth]{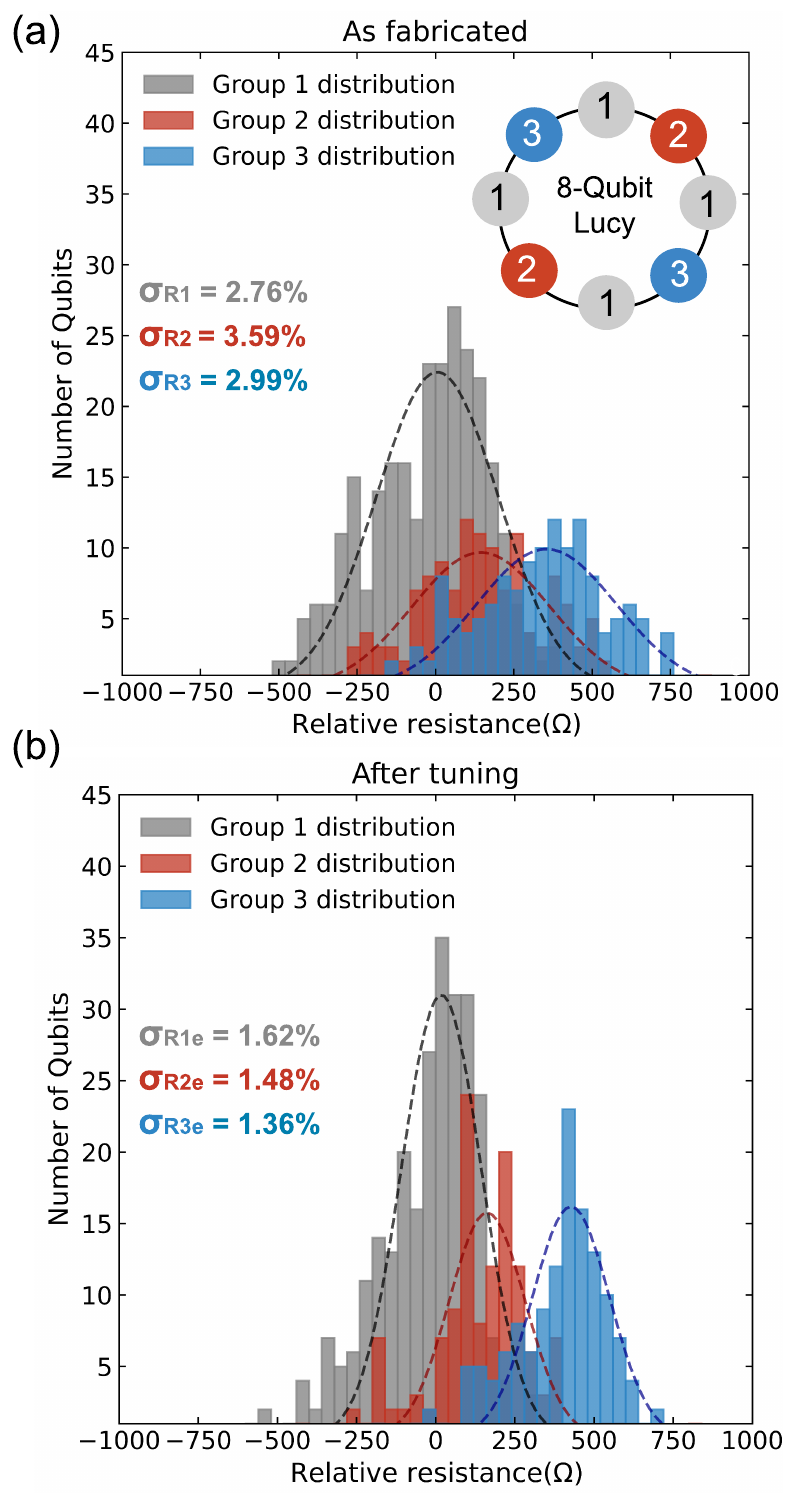}
\caption{\label{fig:wafer}(a) The inset shows the OQC Lucy 8-qubit ring topology with three targetted qubit frequencies (JJ-1 denoted by gray-1, JJ-2 by red-2, and JJ-3 by blue-3). The histograms show room temperature resistance measurements for each targetted junction type as fabricated - i.e., with no electron-beam annealing treatment. The spread for each junction type is noted. (b) shows the histograms of the same junctions following an e-beam tuning treatment. We present the numerical spread for the histograms, with the Gaussian fits serving as a visual guides. The spread of each junction type is reduced in each case. Group 1 shows a numerical spread reduction from $\sigma_{\mathrm{R1}}$ =  2.76\,\% to $\sigma_{\mathrm{R1e}}$= 1.62\,\%, Group 2 shows a reduction from $\sigma_{\mathrm{R2}}$ = 3.59\,\% to $\sigma_{\mathrm{R2e}}$ = 1.48\,\%, and Group 3, $\sigma_{\mathrm{R3}}$ =   2.99\,\% to $\sigma_{\mathrm{R3e}}$ = 1.36\,\%.}
\end{figure}

\begin{figure*}
\includegraphics[width=1\textwidth]{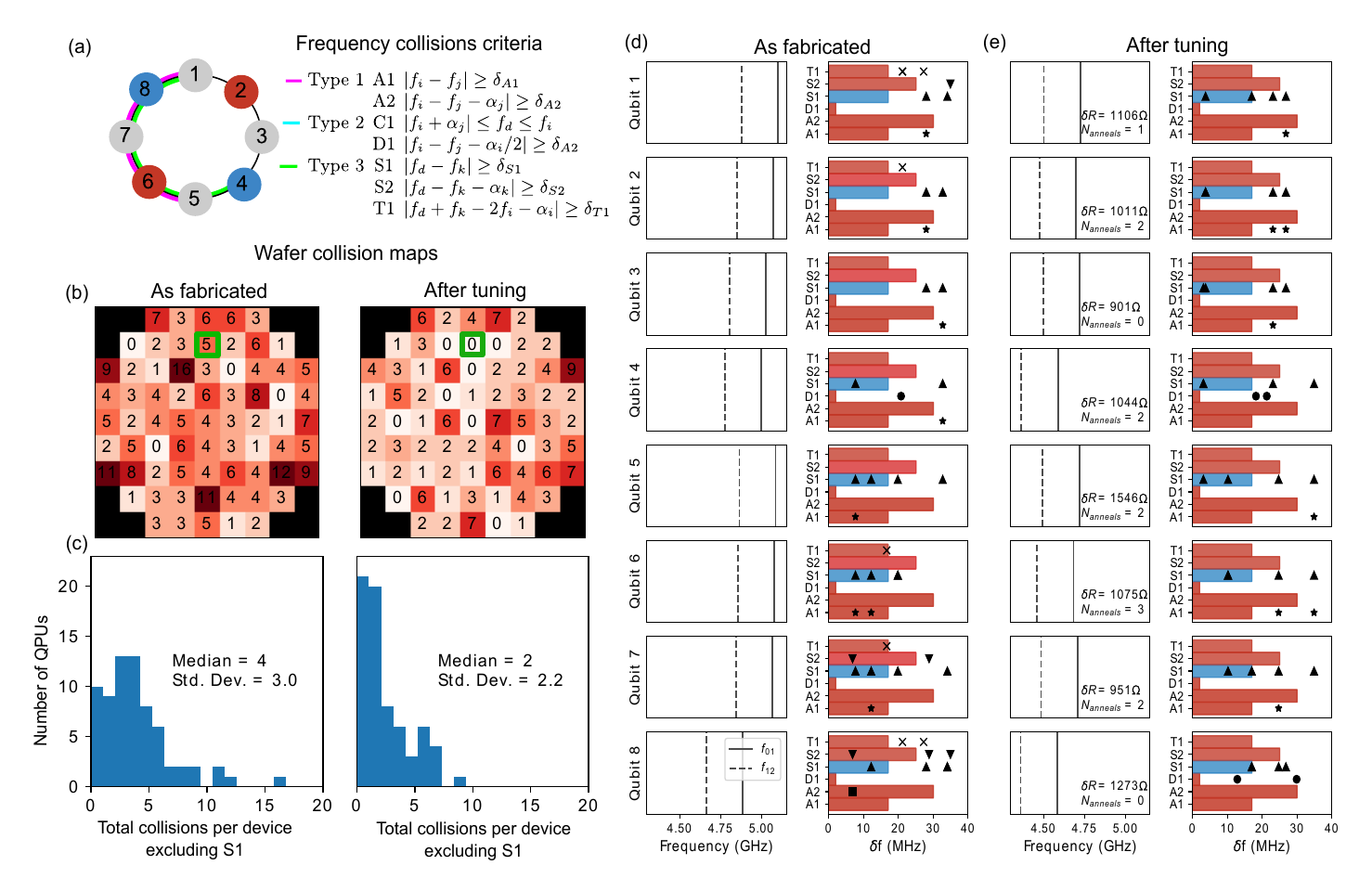}
\caption{\label{fig:collisions}(a) Illustrates an 8-qubit device highlighting the anticipated collisions, based on the frequency collision criteria described in Ref  \cite{Morvan2022}. The criteria comprise of the transmon frequency, $f_{i}$, with its neighbour, $f_{j}$, and next-nearest neighbour, $f_{k}$. The frequency of the applied drive microwave is represented by $f_{d}$, the anharmonicity is described by $\alpha$, and $\delta$ describes the minimum detuning to avoid frequency collisions. (b) Wafer scale colour maps of expected collisions before and after tuning within each 8QPU die. Many QPUs show reduced numbers of expected 2Q collisions, in some cases reduced to zero, such as the 8QPU outlined in green, and which is illustrated in (a). (c) Histograms of total collisions (excluding S1 collisions) per QPU over a full 3 inch wafer as-fabricated (left panel) and post-tuning (right panel). (d) The expected frequency collisions for the QPU shown in (a) and highlighted green in (b) is evaluated by assessing each qubit  before and after tuning. In this context, the first column in (d) illustrates the qubit frequencies for the ground-state (solid line), and excited-state (dashed line), also considered as the C1 criteria.  The second column in (d) provides a corresponding assessment of potential collisions based on the criteria of Ref  \cite{Morvan2022}. Symbols falling within the red bars mean that a frequency collision will occur.  (e) shows the qubit frequencies of the ground-state (solid line), and the excited-state (dashed line) as well as expected collisions post-tuning. }
\end{figure*}

\section{\label{sec:level5}Wafer level spread reduction}

To demonstrate the effectiveness of the e-beam junction tuning technique, we also performed wafer scale resistance tuning on OQC's Lucy-generation 8QPU's  \cite{Lucy8QPU}. The inset in Figure \ref{fig:wafer}a shows a schematic of the 8-qubit lattice, which follows a ring topology, with the qubits operating at 3 distinct frequencies (and thus 3 different designed JJ dimensions) spaced appropriately to avoid any nearest-neighbour collisions between the qubits. We start by fabricating a wafer containing 8QPU dies, followed by room-temperature measurements of the JJ tunnel barrier resistances. Figure \ref{fig:wafer}a shows the resistance distribution of the three JJ dimensions grouped as 1, 2, and 3, along with their corresponding standard deviations ($\sigma_{\mathrm{R}}$).  The normalized cumulative distribution after JJ fabrication is,  $\sigma_{\mathrm{R1,2,3}}$ = 3.11\,\% across 552 JJs on a 3 inch wafer.

After employing three stages of e-beam tuning (following the procedure outlined in  in section \ref{sec:level2}), the cumulative distribution was reduced to $\sigma_{\mathrm{R1,2,3e}}$ = 1.48\,\%, as shown in Figure  \ref{fig:wafer}b.  Note that only JJs with resistances between 80-120\,\% of our median are included in the spread statistics. Values outside of this range are out-of-specification and contribute to our yield statistics as failed JJs. In order to demonstrate that tuning is only performed locally on the targetted JJs,  we plot resistance shift before and after tuning in Figure \ref{fig:S2}c with more details in the Appenidix A. The wafer yield following the three rounds of JJ tuning remains stable, with only a slight decrease of 1.6\,\% from the as-fabricated yield, thus highlighting the reliability of the technique.

\begin{figure*}[t]
\includegraphics[width=0.7\textwidth]{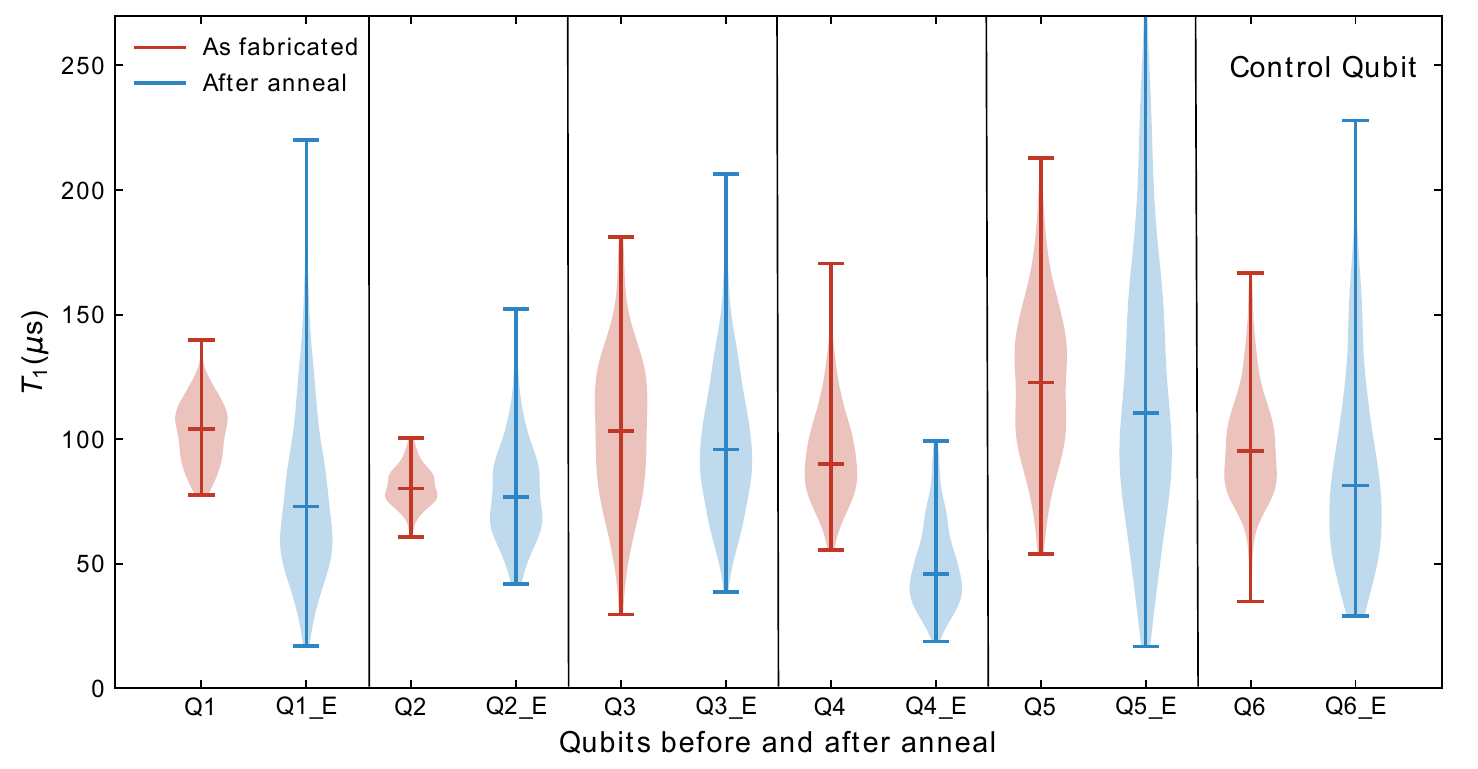}
\caption{\label{fig:coherence} Comparison of $T_{1}$ coherence time measurements for six qubits. The qubit die was measured prior to carrying out the e-beam annealing. Following measurements of $T_{1}$ for each qubit, the die was warmed up to room temperature and unpackaged. Five of the six qubits underwent the e-beam annealing with qubit 6 left as a control qubit. The die was repackaged, recooled, and the coherence metrics were remeasured. The results show that the qubits remain as high-coherence qubits post-anneal.}
\end{figure*}
To determine the viability of e-beam tuned QPU's, we evaluate the likelihood of encountering undesirable collisions based on the qubit frequencies. These collision criteria are categorised into three types as described by Morvan \textit{et al} \cite{Morvan2022} and are defined in Figure \ref{fig:collisions}a.  Type 1 (referred as A1 and A2) corresponds to single-qubit resonances, where the transition frequency of the control qubit is required to be different to that of its nearest neighbour. Type 2 (referred as C1, E1, E2 and D1)  involves the resonance between any adjacent qubits where the frequency gap between the qubits are less than their anharmonicity (straddling regime), as well as to avoid  two-photon transitions (D1). We note that we have excluded E1 and E2, since for the CR gate, qubits are driven at the frequency of their neighbouring qubits, making E1 and E2 equivalent to A1 and A2. Lastly, Type 3 (referred as S1, S2 and T1) signifies a 3-qubit resonance condition where the pulse applied to entangle two qubits can also induce one-photon (S1 and S2) or two-photon (T1) transitions in other neighbouring qubits.  Based on the above criteria, we plot the wafer colour maps of the expected collisions in Figure \ref{fig:collisions}b before and after the tuning procedure, where each square corresponds to an individual 8-qubit device. The distribution of wafer-level collisions, visualized in Figure \ref{fig:collisions}c, demonstrates a reduction in overall collisions after e-beam tuning. Note that S1 collisions are intrinsic to our 8PU lattice design, as the group 1 frequencies are positioned in a manner that inevitably leads to S1 type collisions, hence the S1 criteria were intentionally excluded in (b) and (c). To understand further, we provide a detailed assessment in the Appendix A, which examines the likelihood of collisions when resistance spread is simulated to be reduced to zero.

Figure \ref{fig:collisions}a illustrates one of the 8-qubit device selected from Figure \ref{fig:collisions}b, underscoring the initial collision types based on its qubit frequencies before the tuning process. This particular device exhibited a reduction in collisions, from 5 to 0 after the tuning procedure. Figures \ref{fig:collisions}d and \ref{fig:collisions}e provides an evaluation of the 8 qubits on the various collision metrics before and after the tuning process respectively. In the first columns of both figures, the qubit frequencies for the ground state (solid line) and the excited state (dashed line) are presented to assess the C1 criteria. The C1 criteria evaluates whether the frequency gap between neighbouring qubits falls within the straddling regime. The second columns of  Figure \ref{fig:collisions}d and \ref{fig:collisions}e provide a corresponding assessment of potential collisions in every qubit based on each criteria. Symbols falling within the red bar signify an impending frequency collision.  As expected we observe that the number of S1 collisions (indicated by the blue bar) increase after the tuning process.

\section{\label{sec:level5}Coherence  measurements after e-beam annealing}

The primary focus of this research is on developing the capability to tune the resistances of post-fabricated JJs. This tuning technique is developed to improve qubit frequency allocation, reducing the occurrence of undesirable collisions, and enhancing gate fidelities. This technological advancement relaxes one of the barriers for QPU scaling. However, in addition to frequency allocation, qubit coherence remains important.

To evaluate the impact of our e-beam tuning technique on qubit coherence, we conducted measurements of the qubit relaxation time, $T_{1}$, for six qubits both before and after the annealing process. All JJs that underwent annealing were subjected to specific dose conditions intended to modify their resistance.  Figure \ref{fig:coherence} shows the $T_{1}$ relaxation time measurements for a qubit die before, and after, electron-beam annealing process. Qubits 1, 2, 3, 4, and 5 had the electron-beam annealing procedure performed on them. Qubit 6 was a control qubit, so no e-beam annealing was performed on this qubit. Crucially, the frequencies of the qubits before and after the annealing process displayed no significant variation in their `usual' resistance-frequency relationship.  Figure \ref{fig:coherence} shows the qubits $T_{1}$ times before and after electron-beam annealing. The measurements show that in all cases the qubits retained their high coherence times after the annealing process. On average, a $21\,\mu\mathrm{s}$ reduction in $T_{1}$  was observed, however a reduction of $16\,\mu\mathrm{s}$ also occurred for the control qubit with no e-beam exposure. This
reduction may be attributed to the thermal cycling of the QPU and the removal and subsequent reinstallation of the QPU into its processor package however we cannot rule out the additional processing and handling carried out to perform the electron-beam annealing procedure, such as the application of the charge mitigation layer prior to the annealling process.

\section{\label{sec:level5}Conclusion}

In conclusion, we have demonstrated the ability to locally tune superconducting qubits via application of an electron-beam directly onto, or in the near-vicinity of, the qubit's JJ element. Electron-beam tuning was used to locally tune the Josephson junction resistance (correspondingly tuning its frequency). The sensitivity of the junction resistance shift is characterised with respect to both e-beam current and the proximity to the junction, allowing fine control of the junction resistance to meet specific frequency targets.
We demonstrate a reduction in resistance (frequency) spread of our qubits by locally tuning Josephson junctions that are lower in resistance than that targetted. We have tuned QPUs at wafer scale, containing a total of 552 Josephson junctions. We reduced the resistance spread from an as-fabricated 3.11\,\% to 1.48\,\%, following three e-beam tuning processes. By assessing the expected frequency collisions in our processors we demonstrate that e-beam tuning can be used as a tuning tool to better allocate frequencies in order to reduce frequency collisions and improve expected QPU performance as well as increase the yield of viable QPUs. Finally, we presented results of qubit $T_{1}$ coherence times and establish that our qubits remain high-coherence after they have undergone the e-beam tuning treatment. With further refinement of this technique we expect to further the range of tuning that is possible using this process, thus further reducing Josephson junction spread allowing improved qubit frequency allocation. This  technique can be performed using tools such as an EBL or a SEM, which allow spatial control of the e-beam. Since these tools are readily available in fabrication facilities, this technique is suitable for integrating into Josephson junction fabrication workflows and processes.

\begin{acknowledgments}
We extend our thanks to the entire OQC Team for their contributions to the quantum computing stack, which was instrumental in this work. Special thanks to Dr.~Ilana Wisby, Simon Phillips,  Brian Vlastakis, Jonathan Burnett, and Peter Leek for their review of this manuscript.  We thank Rais Shaikhaidarov and Phil Meeson for their support at the Royal Holloway University of London SuperFab Facility. Our heartfelt appreciation also goes to our investors and users for their unwavering support.

We gratefully acknowledge the indispensable contributions of our team members, whose collective efforts were essential to the success of this publication.
Waqas Ahmad, Aleena Alby, Katy Alexander, Owen Arnold, Ofek Asban, Amir Awawdi, Charlie Bach, Kristian Balog, Luke Batty, Bryn Bell, Ambika Bhatia, Anirban Bose, Rich Bounds, Richard Bowman, Ray Brown, Tina Cannings, Ese Charles-Adeoye, Rodrigo Chaves, Boris Chesca, Philip Clarke, Gioele Consani, Nikki Cooper, Abbie-Rose Curbison, Wassem Dabbas, Charlie Dale, James David, Will David, Esten De Souza, Norbert Deak, Gavin Dold, John Dumbell, Jacob Dunstan, Sam Earnshaw, Caro Ehrman, Hamid El Maazouz, Kajsa Eriksson Rosenqvist, David Foster, Jamie Friel, Amber Glennon, Chelcia Gordon, Mohammad Tasnimul Haque, Anna Harrington, Darren Hayton, Apoorva Hegde, Will Howard, Aamna Irfan, Kaitlyn Jannetta, Chris Joy, Ines Juvan-Beaulieu, Ailsa Keyser, Ilona Kozlowska, Shelley Lam, Andrew Lennard, Alex Lillistone, George Long, Antonio Maieli, Seja Majeed, Martin Millmore, Gerald Mullally, Chris Nicholson, Peter Oakley, Deepali Parti, Richard Pearson, Lee Peters, Viviana Pol-Waterston, Charles Prochazka, Daryl Rees, Steven Reeve, Ben Rogers, Russell Rundle, Michelle Scott, Minsu Seo, Boris Shteynas, Connor Smith, Mark Stainer, Gillian Steele, Sandy Strain, Atsushi Sugiura, Nicola Taylor, Habib Ullah, Travers Ward, Harry Waring, Ryan Wesley, Ken Westra, James Wills, Tom Winchester, and Marco Zaratiegui.

\end{acknowledgments}

\bibliography{EBLA-CDS-arxiv-submit}

\newpage

\appendix

\section{\label{sec:level5}Localised effect of e-beam tuning}
\begin{figure}
\includegraphics[width=0.5\textwidth]{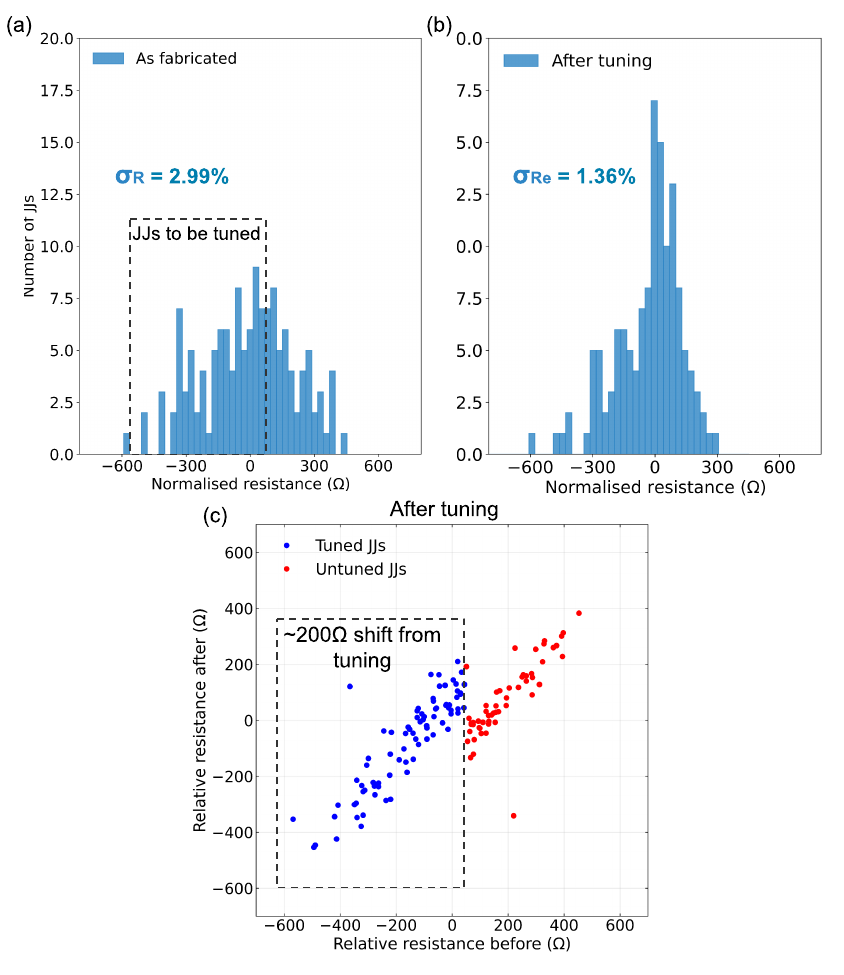}
\caption{\label{fig:S2}(a) Histogram of room temperature resistance measurements for JJs with the same designed dimensions as-fabricated - i.e., with no e-beam tuning. The dashed black box highlights the JJs that will undergo e-beam tuning. (b) Histogram of the same JJs after e-beam tuning. (c) plots the comparison between the relative resistances before and after e-beam tuning from (a) and (b) respectively. The JJs that underwent e-beam tuning  show a larger resistance shift of approximately 200~$\Omega$ compared to the JJs that did not undergo tuning.}
\end{figure}

In order to empirically demonstrate the local tuning capabilities of the e-beam on targeted JJs, we performed analysis on the resistance shift before and after tuning. Figure \ref{fig:S2}a plots the histogram of the fabricated JJ resistances. The standard deviation is  $\sigma_{\mathrm{R}}$ =  2.99\,\%. Figure \ref{fig:S2}b shows the resistance distribution following e-beam tuning, which effectively reduced the standard deviation to $\sigma_{\mathrm{Re}}$ =  1.36\,\%. To illustrate the precision of this tuning, we focused on the part of the  resistance distribution with a relative resistance of 50~$\Omega$ or below (represented by the dashed box in figure \ref{fig:S2}a). Figure \ref{fig:S2}c plots the resistance values before and after the tuning process. Notably, we observe that the tuned JJs were subjected to an additional 200~$\Omega$ increase in resistance compared to their untuned counterparts. We applied a fixed dose to all the tuned JJs, which resulted in a consistent resistance shift. This observation highlights that the e-beam locally accelerated the ageing process of these JJs, differentiating them from the untuned JJs.

\section{\label{sec:level5}Assessment of S1 collisions in a 8QPU lattice}

\begin{figure}
\includegraphics[width=0.4\textwidth]{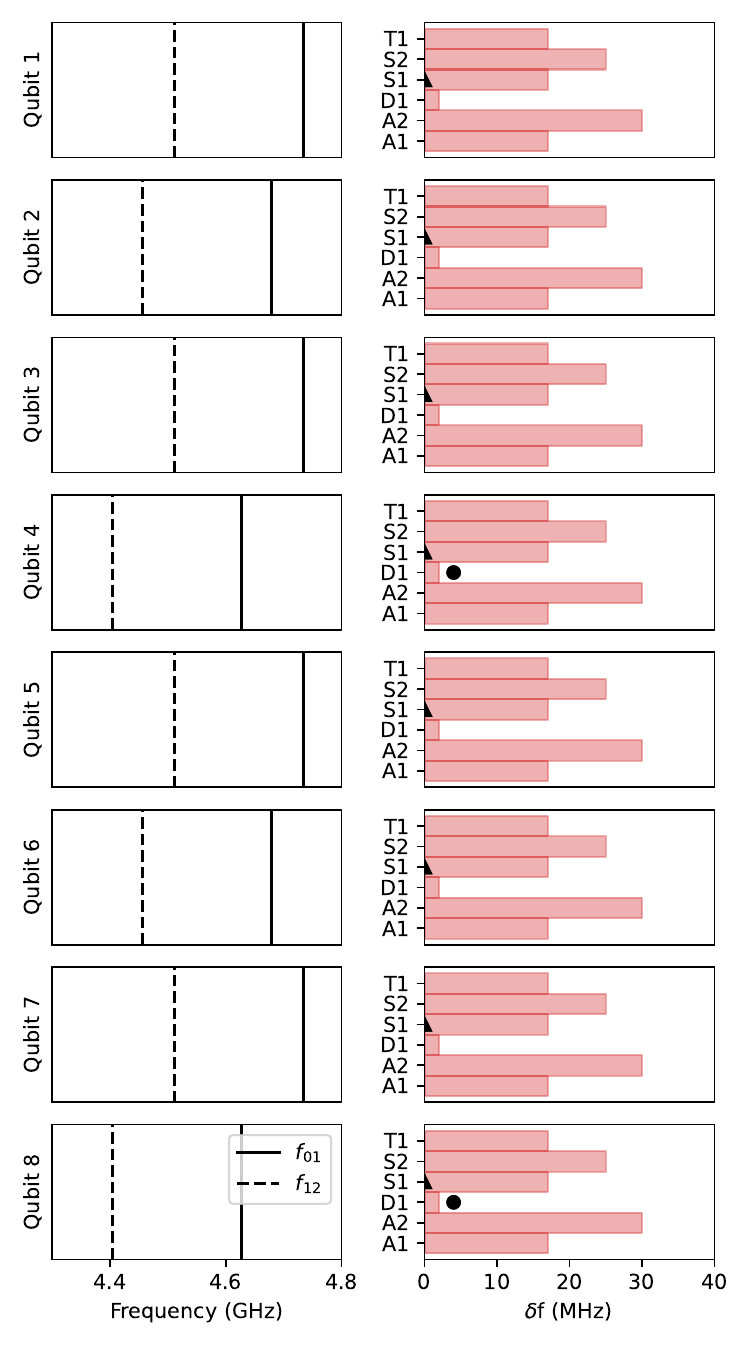}
\caption{\label{fig:S3} Simulated assessment of the potential frequency collisions in the 8-qubit lattice, assuming zero spread between qubit frequencies. The first column corresponds to the qubit frequencies of the ground-state (solid line), and excited-state (dashed line). The second column shows a corresponding assessment of potential collisions. Symbols falling within the red bars mean that a frequency collision will occur. This figure shows this QPU design is prone to S1 collisions when frequency targetting is perfect (i.e., zero spread). }
\end{figure}

In this analysis we motivate the decision to exclude S1 collisions in the assessment of collision reductions in the main manuscript. As our 8QPU topology has only 3 targetted frequencies we are susceptible to this collision type. Because of this, as we reduce our frequency spread we expect the S1 collisions to become more frequent in our QPUs. Indeed, we did observe an increase in S1 collision as we reduced the spread of our JJs (see S1 collisions in fig \ref{fig:collisions}e). We simulate the QPU collision metrics as if the as-fabricated JJs exhibited no spread i.e., the resistances were exactly as designed. Figure \ref{fig:S3}  illustrates the collision outcomes corresponding to zero spread.  We observe that all clashes are absent apart from S1, which shows a maximum magnitude of occurrence.  This is attributed to group 1 qubits and their location in our 8QPU ring topology. An alternative QPU JJ targetting scheme would mitigate these S1 collisions.

\end{document}